\documentclass[%
 reprint,
 floatfix,
longbibliography,
 amsmath,amssymb,
 aps, physrev,
]{revtex4-2}

\usepackage{tabularx}
\usepackage{graphicx}
\usepackage{dcolumn}
\usepackage{bm}
\usepackage{braket}
\usepackage{appendix}
\usepackage{natbib}
\usepackage{amsmath}
\usepackage{hyperref}
\usepackage{booktabs}

\usepackage{dsfont}
\usepackage{verbatim}

\begin{document}


\title{Sub-resonant wideband superconducting Purcell filters}

\author{Basil M. Smitham}
\email{bsmitham@princeton.edu}
\affiliation{Department of Electrical and Computer Engineering, Princeton University, Princeton, New Jersey 08540, USA}
\author{Jeronimo G. C. Martinez}
\affiliation{Department of Electrical and Computer Engineering, Princeton University, Princeton, New Jersey 08540, USA}
\author{Christie S. Chiu}
\affiliation{Department of Electrical and Computer Engineering, Princeton University, Princeton, New Jersey 08540, USA}
\author{Andrew A. Houck}
\email{aahouck@princeton.edu}
\affiliation{Department of Electrical and Computer Engineering, Princeton University, Princeton, New Jersey 08540, USA}

\date{\today}

\begin{abstract}

In superconducting quantum devices, Purcell filters protect qubit information from decaying into external lines by reducing external coupling at qubit frequencies while maintaining it at readout frequencies. Here, we introduce and demonstrate a novel Purcell filter design that places the readout resonator frequencies in a ``linewidth plateau" below the filter's first resonant mode. This approach, based on direct admittance engineering, can simultaneously achieve strong qubit protection and nearly constant external coupling across a wide readout bandwidth---addressing the traditional tradeoff between these properties. We first present a lumped-element analysis of our filters. We then experimentally demonstrate a compact on-chip linewidth-plateau filter, coupled to four resonators across its approximately $1$ GHz readout band. We compare the measured linewidths to numerical predictions, and show how the filter protects a frequency-tunable transmon qubit from external decay. We envision that our flexible design paradigm will aid in efforts to create multiplexed readout architectures for superconducting quantum circuits, with well-controlled external couplings.

\end{abstract}

\maketitle

\section{Introduction} \label{sec:Introduction}

Superconducting qubits must be connected to external circuits for control and measurement. However, these lines serve as additional qubit decay channels. To prevent unwanted qubit decay through measurement circuitry, one frequently deploys a Purcell filter, which uses the Purcell effect \cite{purcell_e_m_proceedings_1946} to reduce external coupling at the qubit transition frequency $\omega_q$, while maintaining the necessary high external coupling at the readout frequency $\omega_r$.

Most Purcell filters are adaptations of existing $50$ $\Omega$ microwave power transfer filters \cite{david_m_pozar_microwave_2012}---where the qubit transition is placed in the filter's stopband and the readout transition in the filter's passband. Examples in quasi-planar superconducting architectures include quarter-wave bandstop filters \cite{reed_fast_2010}, resonator bandpass filters \cite{sete_quantum_2015, jeffrey_fast_2014, walter_rapid_2017, heinsoo_rapid_2018}, and higher order resonant bandpass filters \cite{bronn_broadband_2015, cleland_mechanical_2019, yan_broadband_2023, park_characterization_2024}. In non-planar architectures, these filters can take the forms of three-dimensional waveguides \cite{narla_robust_2016} and three-dimensional resonators \cite{wang_cavity_2019}. Not all Purcell filters are based on power transfer designs. In particular, there exist interference Purcell filters, in which multiple decay paths interfere to suppress qubit emission \cite{houck_controlling_2008, sunada_fast_2022, spring_fast_2024, yen_interferometric_2024}.

To perform multiplexed readout of multiple qubits using a single Purcell filter, we would ideally want a large frequency bandwidth for the readout resonators and substantial decay protection at the qubit frequencies. Single-pole resonant filters possess an inherent tradeoff between resonator bandwidth and qubit protection \cite{sete_quantum_2015}. Multi-pole resonant filters with additional circuit elements can ameliorate this issue, but they encounter additional problems when operated in the standard ``single-sided" configuration---where the filter is strongly coupled to an output port and only weakly coupled to an input port, such that readout photons decay preferentially into the output detector \cite{heinsoo_rapid_2018}. In this single-sided regime, because the filter is no longer matched to two $50$ $\Omega$ impedances on either side, external readout coupling can vary widely over the readout bandwidth \cite{yan_broadband_2023}. Variation in external linewidth inhibits our ability to engineer high-fidelity qubit readout.

In this work, we introduce a ``linewidth-plateau" design methodology for broadband Purcell filters that simultaneously provides strong qubit protection and precise control of readout resonator external couplings. We adapt the standard multi-pole high-pass filter with alternating series capacitance and shunt inductance segments, but operate it in a sub-resonant regime, where both qubit and resonator modes fall below the filter's first resonant mode. Rather than engineering conventional passbands and stopbands, we instead directly shape the external coupling spectrum---to be approximately constant over the readout bandwidth, enabling more precise targeting of readout resonator linewidths. Below the plateau, the filter's high-pass configuration inherently ensures qubit protection.

\begin{figure*}
    \centering
    \includegraphics[width=.9\linewidth]{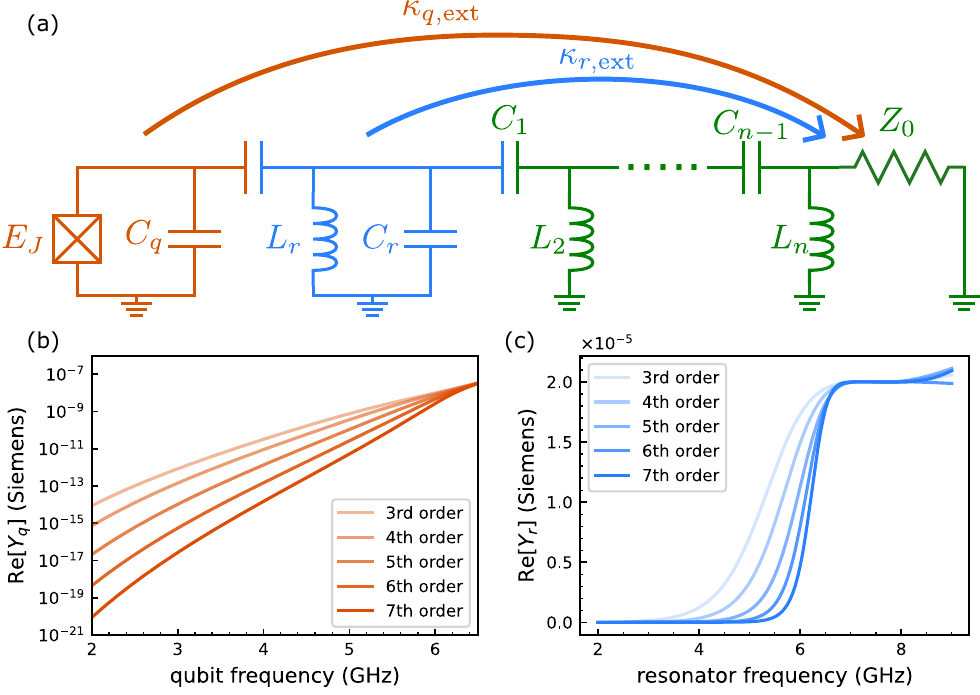}
    \caption{Lumped-element linewidth-plateau filters. Plots of example filters are made with the parameters shown in Table \ref{table:FilterParams}. (a) Lumped-element schematic of external decay of a transmon qubit (orange) and a readout resonator (blue) through a filter (green). The filter consists of alternating series capacitance and shunt inductance. (b) Real part of admittance (proportional to external decay) of the qubit as a function of qubit frequency, tuned by varying the Josephson junction inductance, with the resonator frequency fixed at $7.1$ GHz. By adding filter segments, the decay decreases more rapidly as you detune the qubits below the resonator frequencies. (c) Resonator admittance/decay profile through the filter. The filters are numerically designed to have constant decay profile at the linewidth plateau between $7$ and $8$ GHz, which lies below the filter's resonant modes.}
    \label{fig:LumpedFilter}
\end{figure*}

We first illustrate the theoretical basis of our approach with a simplified model: a transmon qubit and its readout resonator coupled to an external transmission line through a lumped-element capacitive/inductive filter circuit. We numerically demonstrate how to optimize the filter designs to create linewidth plateaus, where readout resonator external decay $\kappa_{r,{\text{ext}}}$ varies slowly as a function of resonator frequency. By increasing the number of capacitive and inductive filter elements, we raise the order of the filter, boosting suppression of external qubit decay at frequencies below the plateau.

\begin{table} 
\begin{tabular}{|c|c | c | c | c | c|}
    \hline
    filter order & 3rd & 4th & 5th & 6th & 7th  \\
     \hline
 $C_1$ (fF) & 7.92 & 9.13 & 7.40 & 8.32 &  8.63  \\
 $L_2$ (nH) & 2.61 & 1.64 & 1.97 & 1.38 &  1.81  \\
 $C_3$ (fF) & 297 & 289 & 198 & 257 &  136  \\
 $L_4$ (nH) &  & 1.70 & 1.02 & .632 &  1.38  \\
 $C_5$ (fF) &  &  & 640 & 398 &  154  \\
 $L_6$ (nH) &  &  &  & 1.43 &  1.20  \\
 $C_7$ (fF) &  &  &  &  &  424  \\
\hline
\end{tabular}
    \caption{Parameters for the linewidth-plateau filters of varying orders used to make the admittance plots in Fig. \ref{fig:LumpedFilter}.}
    \label{table:FilterParams}
\end{table}

Next, we present an experimental design and realization of a superconducting linewidth-plateau filter device. The filter, which consists of four alternating series capacitive and shunt inductive segments, mediates coupling between coplanar waveguide (CPW) resonators and a 50 $\Omega$ external line. The resonator frequencies lie within the theoretically-predicted $7$ to $8$ GHz linewidth plateau, and the lowest-frequency resonator is coupled to a tunable-frequency transmon qubit. We measure the resonators' linewidths and compare them to numerically-calculated predictions, noting the deviation resulting from impedance mismatch in our packaging. We then measure the qubit's relaxation time $T_1$ as a function of its resonance frequency, demonstrating that the filter protects the qubit from Purcell decay.

\section{Sub-resonant Purcell filters} \label{sec:SubResPurcellFilters}

Fig. \ref{fig:LumpedFilter} gives an outline of our design methodology for linewidth-plateau filters, within a lumped-element circuit framework. We use admittance engineering to shape the decay profiles of the qubit and readout resonator modes, with their transitions located far below the first resonant mode of the filter. In particular, our design makes the resonator's external linewidth approximately constant as a function of frequency within a specified bandwidth.

Fig. \ref{fig:LumpedFilter}(a) shows the circuit schematic of a coupled transmon qubit (orange), readout resonator (blue), and linewidth-plateau filter (green). The filter here consists of a variable number of alternating series capacitors and shunt inductors. We emphasize two key design parameters determined by the filter: the external coupling of the readout resonator $\kappa_{r,{\text{ext}}}$ and the unwanted external decay rate of the qubit $\kappa_{q,{\text{ext}}}$. As in \cite{houck_controlling_2008}, for weakly anharmonic modes, these quantities are proportional to the real part of admittance:
\begin{align}\label{eq:kappa_i}
\kappa_{i{,\text{ext}}}(\omega_i) = \frac{ \text{Re} [Y_i(\omega_i)]}{C_i(\omega_i)}
\end{align}
where the subscript $i$ denotes either the resonator or the transmon qubit, and $\omega_i$ refers to their characteristic resonant frequencies. The real part of admittance $\text{Re}[Y_i]$ (which quantifies the loss of the mode through the output line) and the total mode capacitance $C_i$ are computed with respect to ground from either the resonator or qubit circuit nodes.

In order to probe external linewidth at different resonator and qubit frequencies, we vary the resonator inductance and the qubit (Josephson junction) inductance, while leaving all capacitances constant. Then, the frequency-variation of external coupling (Eq. \ref{eq:kappa_i}) can be determined solely by $\text{Re}[Y_i(\omega_i)]$.

In Figs. \ref{fig:LumpedFilter}(b) and \ref{fig:LumpedFilter}(c) we plot $\text{Re}[Y_q(\omega_q)]$ and $\text{Re}[Y_r(\omega_r)]$ as a function of qubit and resonator frequencies $\omega_q/(2 \pi)$ and $\omega_r/(2 \pi)$, respectively---for linewidth-plateau filters with varying number of circuit elements, whose parameters are given in Table \ref{table:FilterParams}. As shown in Fig. \ref{fig:LumpedFilter}(c), the design is numerically optimized to have a roughly constant $\text{Re}[Y_r]$ of $2 \times 10^{-5}$ Siemens in the wide frequency range of $7$ to $8$ GHz, giving the readout resonator an approximately constant external decay rate $\kappa_{r,{\text{ext}}} \propto \text{Re}[Y_r(\omega_r)]$ as a function of resonator frequency.

These lumped-element models can be studied analytically, in order to gain insight into the origins of the sub-resonant linewidth plateau. As discussed in Appendix \ref{app:LinewidthPlateauFilters}, with this filter structure, the resonator's real part of admittance rate can be written as:
\begin{align}
\text{Re} [ Y_r(\omega)] &=
            \frac{\omega^{2N}}{P_{N}(\omega^2)}
\end{align}
where $N$ is the number of circuit elements in the filter, and $P_{N}(\omega^2)$ is a polynomial of order $N$ in terms of $\omega^2$ with nonzero constant term. Complex values of $\omega$ near the real axis where $P_{N}(\omega^2) \rightarrow 0$ correspond to the filter's modes. We engineer the real frequency component of any such pole $\text{Re}(\omega_p)$ to be much higher than the operating qubit and resonator frequencies. Then, below the first resonance, at low frequencies, the real part of admittance scales as: $\text{Re} [ Y_r(\omega)] \propto \omega^{2N}$. Near poles, this real part of admittance will have a peak. However, leading up to the pole, different shapes of the real admittance response can be engineered. In particular, the real part of admittance can be designed to have a roughly flat response over a certain frequency band by interfering terms of different orders in the denominator polynomial $P_{N}(\omega^2)$, forming the linewidth plateau.

In Fig. \ref{fig:LumpedFilter}(b) we examine qubit Purcell decay protection at frequencies below the resonator linewidth plateau. To do this, we fix the readout resonator frequency at $7.1$ GHz and plot the real part of the qubit's admittance $\text{Re}[Y_q(\omega_q)]$ versus the qubit frequency $\omega_q/(2 \pi)$. As the number of filter circuit elements increases, so does the sharpness with which $\kappa_{q,{\text{ext}}}(\omega_q) \propto Y_q(\omega_q)$ declines, as the qubit frequency is detuned below the frequencies of the resonator and linewidth plateau---so qubit protection can be improved while maintaining constant external readout coupling. We note that the linewidth-plateau design only functions when qubits lie below the readout resonator frequency band.

\section{Linewidth-plateau filter device} \label{sec:LinewidthPlateauFilterDevices}

\begin{figure}
    \centering
    \includegraphics[width=1\linewidth]{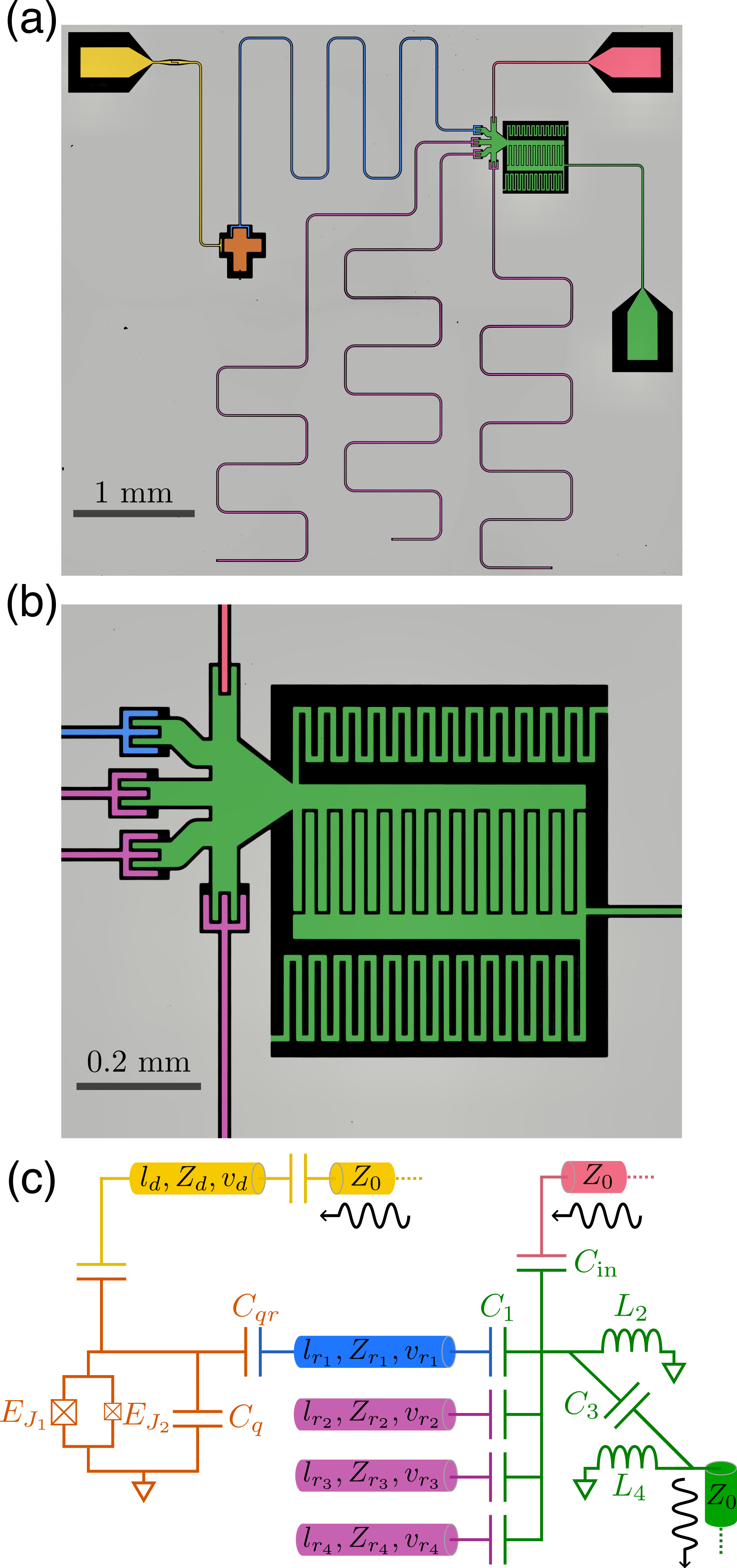}
    \caption{Device image and circuit model. (a) Device image showing the filter and output line (green), input line (coral pink), resonators (blue and purple), qubit (orange), and qubit drive line (yellow). The drive line contains a high-frequency resonator filter, which reduces external coupling at the qubit frequency---allowing us to couple the qubit to the external drive with a larger, closer capacitor. (b) Zoomed-in image of the 4th order filter. (c) Circuit model approximation of the device, including coplanar waveguide segments of length $l$, characteristic impedance $Z$, and phase velocity $v$. The input and output lines ideally terminate in a  characteristic impedance of $Z_0 = 50$ $\Omega$.}
\label{fig:DeviceLayout}
\end{figure}

We now detail the design, fabrication, and measurement of a device with a linewidth-plateau Purcell filter. Though the filter is based initially on a lumped-element model, the design process must account for additional parasitic couplings, ensuring that they do not degrade the desired behavior of the filter.

Fig. \ref{fig:DeviceLayout}(a) is an optical image of the device layout, Fig. \ref{fig:DeviceLayout}(b) is a zoom-in on the Purcell filter, and Fig. \ref{fig:DeviceLayout}(c) gives an approximate circuit schematic. The device consists of four resonators (blue and purple) coupled to an output port via a 4th order sub-resonant linewidth-plateau filter (green). From the perspective of each resonator, the filter consists of two series capacitors alternating with two shunt inductors coupled to a $50$ $\Omega$ output port. In addition to its strongly-coupled output port, the device also is weakly coupled to an input port (coral pink).

One of the readout resonators (blue) is further coupled to a frequency-tunable transmon qubit (orange), which has its own AC charge drive line (yellow). This charge line employs a high-frequency resonator through which the qubit is off-resonantly driven, reducing external coupling and enabling a larger, closer coupling capacitor. To tune the qubit frequency, we thread flux through the qubit's SQUID loop with an external magnet. The qubit's $T_1$ decay time is measured and used to probe the device's Purcell filtering properties.

To design the filter, we perform full-wave electromagnetic (EM) simulations using the Cadence AWR AXIEM (method of moments) solver. We first conduct high-frequency EM analysis of a cutout region around and including the filter, placing electromagnetic ports between the center pin and ground for each of the resonators, as well as on the input and output lines. Across the input and output ports, $50$ $\Omega$ resistors were inserted to mimic the dissipative external environment.

The external decay rates were then calculated from Eq. \ref{eq:kappa_i} by constructing an effective LCR model from the perspective of each resonator port---near the resonance frequency. We probe the system over a wide frequency range by varying the simulated CPW half-wave resonator length. Note that unlike in the procedure from Fig. \ref{fig:LumpedFilter}, the effective capacitance $C_r(\omega_r)$ now also varies as a function of resonator frequency, as in \cite{goppl_coplanar_2008,wisbey_new_2014} (described with more detail in Appendix \ref{sec:EffectiveCapacitanceCPW}). Recall that $\kappa_{r,\text{ext}}(\omega_r) = \text{Re} [Y_r(\omega_r)]/ C_r(\omega_r)$, so that the external linewidth will now be the quotient of two frequency-dependent quantities. The filter design determines $\text{Re}[Y_r(\omega_r)]$, and we iterate over the design space to compensate for the frequency-domain effects of $C_r(\omega_r)$ and maintain the linewidth plateau.

\begin{figure}
    \centering
    \includegraphics[width=1\linewidth]{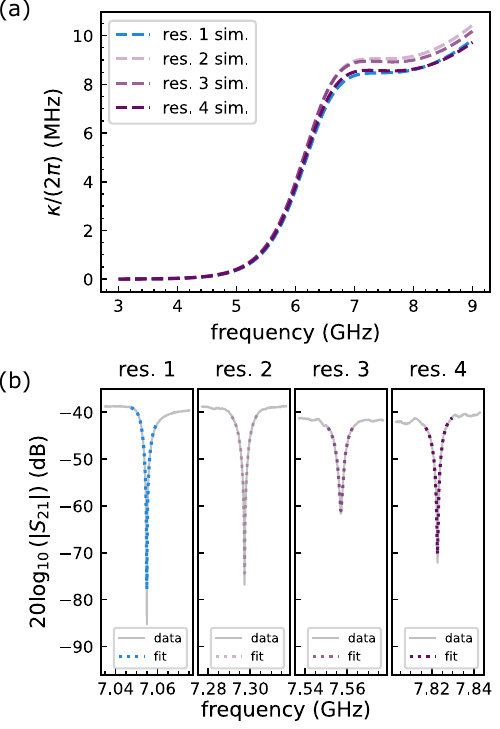}
    \caption{Simulations and measurements of resonator linewidths. (a) Electromagnetic full-wave method of moments simulation of resonator linewidth $\kappa/(2\pi)$, for the filter shown in Fig. \ref{fig:DeviceLayout}. The resonance frequency is tuned by varying the effective length of each of the four resonators. The external linewidths plateau to values of $8$ to $9$ MHz within the frequency range of $7$ to $8$ GHz, for each resonator simulation. (b) $S_{21}$ transmission measurements of the device with four coupled resonators in the frequency range of the predicted linewidth plateau. A circle-fitting procedure is performed to extract the linewidth of each resonator, with the results shown in Table \ref{table:ResFreq}.}
    \label{fig:LinewidthPlateauDistributed}
\end{figure}

The final linewidth design simulations are shown in Fig. \ref{fig:LinewidthPlateauDistributed}(a). Despite the distributed nature of the simulations, we are still able to achieve a good approximation of the target linewidth-plateau profile for the four resonators in the $7$-$8$ GHz range---demonstrating that the linewidth-plateau paradigm does not break down for realistic device layouts. We do observe some overall variation between the simulated decay rates for different resonators, resulting from position-dependent coupling effects. To equalize the external decay rates for each resonator in future designs, we could adjust the coupling capacitances to compensate for these spatial variations.

In Fig. \ref{fig:LinewidthPlateauDistributed}(b) we show the measured transmission profile of the filter around the four resonators, whose frequencies lie between $7.05$ and $7.82$ GHz. The measured linewidths were extracted from a circle fitting procedure (Appendix \ref{sec:CircleFitting}), and range from $7.64$ to $10.65$ MHz, with deviations from the target $8$-$9$ MHz range (and from ideal circular $S_{21}$ profiles) likely resulting from impedance mismatches in the packaging and external lines. The measured resonance frequencies and linewidths are shown in Table \ref{table:ResFreq}. A more detailed discussion of the device's scattering measurements is given in Appendix \ref{sec:ScatteringAndFitting}.

\begin{table}[ht]
    \centering
    \setlength{\tabcolsep}{12pt}
    \begin{tabular}{ccc} 
        \toprule
        & \textbf{Frequency (GHz)} & \textbf{$\boldsymbol{\kappa/(2\pi)}$ (MHz)} \\
        \midrule
        \textbf{Res. 1} & 7.054 & $7.65 \pm 0.02$ \\
        \textbf{Res. 2} & 7.297 & $10.65 \pm 0.36$ \\
        \textbf{Res. 3} & 7.557 & $10.00 \pm 0.40$ \\
        \textbf{Res. 4} & 7.823 & $9.05 \pm 0.27$ \\
        \bottomrule
    \end{tabular}
    \caption{Fitted values of frequencies and linewidths of the resonators coupled to the filter. Qubit frequency errors are below the level of $0.1$ MHz.}
    \label{table:ResFreq}
\end{table}

\section{Qubit decay protection} \label{sec:QubitDecayProtection}

\begin{figure}
    \centering
    \includegraphics[width=1\linewidth]{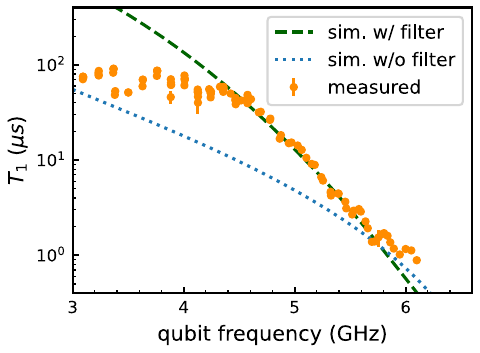}
    \caption{$T_1$ measurements and simulations of the tunable-frequency transmon qubit. The measured $T_1$ values (orange) agree well with the electromagnetically simulated values (green) in the qubit frequency range where the qubit's lifetime is externally limited, before it levels out to its intrinsic limit at low frequencies. Purcell decay protection is demonstrated when compared to a simulation of the qubit's decay rate without the filter (blue).}
    \label{fig:T1Simulation}
\end{figure}

Next, we model and measure the effects of the Purcell filter on qubit relaxation time. Note that the qubit was coupled to resonator $1$. In the low temperature and weakly anharmonic limit of the transmon, the qubit's external decay rate can be written as $\kappa_{q,\text{ext}}(\omega_q) = \text{Re} [Y_q(\omega_q)]/ C_q$, with the external relaxation time limit being the reciprocal $T_{1\text{,ext}} = 1/\kappa_{q,\text{ext}}$. We then calculate $C_q$ and $\text{Re} [Y_q(\omega)]$ through electromagnetic simulation, and measure the qubit $T_1$ as a function of frequency (using the standard dispersive readout scheme) \cite{blais_cavity_2004, wallraff_strong_2004, koch_charge-insensitive_2007}. Both the measured and predicted relaxation times are plotted in Fig. \ref{fig:T1Simulation}. 

We note that this particular filter was designed to have relatively weak qubit protection above $4$ GHz, such that we could observe the transition from externally-limited to internally-limited $T_1$ as the qubit was tuned lower in frequency. As seen in the Figure, there is good agreement between the measured and simulated $T_1$ at externally-limited qubit frequencies. In addition, we see that the filter provides Purcell protection, by comparing the measured $T_1$'s to the external $T_1$ limit without the Purcell filter. The filter-less simulation is a full-wave EM calculation where the qubit couples to the input and output lines through a coplanar waveguide readout resonator (without a Purcell filter) whose external linewidth $\kappa_r$ is designed to match the one measured in the filter device. To achieve better Purcell protection, the filter could be modified with additional capacitive and inductive segments, as laid out in Fig. \ref{fig:LumpedFilter}.

\section{Conclusion}

In this work, we have introduced a new ``linewidth-plateau" methodology for the design of sub-resonant superconducting Purcell filters. These filters can have compact form, high qubit protection from external decay, and slowly-varying readout resonator external coupling over a wide bandwidth. We presented the lumped-element intuition behind these filters in Section \ref{sec:SubResPurcellFilters}, illustrated the design and measurement of the linewidth-plateau properties of a device in Section \ref{sec:LinewidthPlateauFilterDevices}, and demonstrated qubit Purcell protection in Section \ref{sec:QubitDecayProtection}.

Future research could utilize chip package engineering to minimize impedance mismatches, enhancing the flatness of the resonator linewidth plateau. In addition, filters of higher orders (more capacitive and inductive segments) could be tested to observe if external decay is reduced, as predicted in this work. Alternate filter designs could also be explored, such as three-dimensional or nanomechanical architectures. Furthermore, investigations of multiplexed readout could be performed to test the scalability of these designs.

\section{Acknowledgments}

The authors would like to thank Anjali Premkumar for reviewing the manuscript.

Funding for this project was provided by the NSF (grant no. PHY-1607160) and the ARO MURI program (W911NF-15-1-0397). J.G.C.M. acknowledges additional support from the NSF GRFP (DGE-2039656). The device was fabricated in the Princeton University Micro/Nanofabrication Center (MNFC).

Princeton University Professor Andrew Houck is also a consultant for Quantum Circuits Incorporated (QCI). Due to his income from QCI, Princeton University has a management plan in place to mitigate a potential conflict of interest that could affect the design, conduct and reporting of this research.

\appendix

\section{Device, measurement, and simulation}

The superconducting device consists of a $200 \ \text{nm}$-thick tantalum film and aluminum/aluminum oxide Josephson junctions on a $7 \ \text{mm}$ by $7 \ \text{mm}$ by $530 \mu m$ c-plane sapphire substrate. Photolithography was used to define the device design, which was then patterned with a fluorine dry etch. Electron-beam lithography and double-angle evaporation were then employed to define the qubit's aluminum/aluminum oxide/aluminum Josephson junctions. Next, devices were wirebonded to a printed circuit board (PCB) with $50$ $\Omega$ external traces with SMA connectors leading in and out of the device, for input and output lines. Additional wirebonds were placed across the on-chip coplanar lines. A magnetic coil was attached to the rear of the device package, in order to apply a flux bias to the SQUID loop of the transmon qubit.

The devices were mounted on the base plate of a Bluefors LD Dilution Refrigerator, and cooled to a base temperature of $12$ mK. The input lines were outfitted with microwave attenuators and the output line with a high-mobility-electron transistor (HEMT) amplifier.

Frequency domain measurements of the device were carried out with a Keysight PNA-X, and time-domain measurements with the QICK system on a chip (SoC) \cite{stefanazzi_qick_2022}.

Theoretical calculations were performed in the Cadence AWR Microwave Office environment, and the requisite electromagnetic simulations were carried out using the Cadence AWR Axiem electromagnetic solver. AWR Axiem is an open-box method-of-moments microwave simulator for quasi-planar structures. To obtain an accurate estimate of the qubit and resonator decay rates, ``indirect" calculations of real parts of admittance were performed. In the case of the transmon qubit, if port $q$ lies across the qubit's SQUID loop and ports $\{j\}$ are across connections to the external lines (modeled as series resistors with impedance $Z_0 = 50$), then we can estimate this indirect admittance as: $\text{Re}[Y_{q}](\omega_q) \approx \sum_j |Y_{qj}(\omega_q)|^2 Z_0$ \cite{sunada_fast_2022}. The resulting calculation is more accurate in the presence of spurious loss and more numerically stable than the direct computation of $\text{Re}[Y_{q}(\omega_q)]$.

\section{Calculating decay rates}

\begin{figure}
    \centering
    \includegraphics[width=1\linewidth]{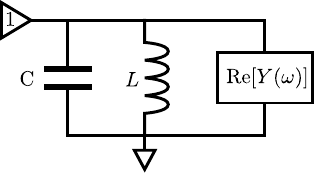}
    \caption{RLC Circuit used to model lossy harmonic (or approximately harmonic) modes. By performing an electromagnetic port simulation, a device's admittance function $Y(\omega)$ may be obtained. From the admittance function's properties near resonance ($\text{Im}[Y(\omega)] = 0$), an effective mode capacitance $C$ and inductance $L$ can be obtained. By evaluating the lossy conductance $1/R = \text{Re}[Y(\omega)]$ at the resonance frequency $\omega_r$, an effective resistance can be obtained.}
    \label{fig:PortRLC}
\end{figure}

To estimate the decay rates (or linewidths) of superconducting resonators, we construct effective parallel LCR models of the resonators, using simulations where a differential electromagnetic port is placed on the resonator (across a region of nonzero voltage at resonance). This setup is shown in Fig. \ref{fig:PortRLC}. We search for the (complex) natural oscillation frequencies where $Y(\omega) = 0$, where $\omega = \omega_r - i \kappa_r/2 $. Here, $\omega_r$ is the resonance frequency and $\kappa_r$ is the resonant linewidth.

However, most commercial electromagnetic simulators only measure port parameters at real frequencies, so methods of network synthesis \cite{newcomb_linear_1966} must be employed to estimate the complex-frequency zeros of admittance. We use the lossy Foster method to approximate the admittance function near resonance with an LCR model \cite{foster_reactance_1924, nigg_black-box_2012}, in order to estimate the location of the zero. This technique assumes that near the real resonance frequency $\omega_r$, admittance can be written as:
\begin{align}\label{YLinear}
    Y(\omega) \approx 2 i C_r \left(\omega - \omega_r \right) + \text{Re}[Y(\omega_r)]
\end{align}
where $C_r$ is the mode's effective capacitance.

Under this approximation, the zero of admittance occurs at the complex frequency of:
\begin{align}
    \omega & =  \omega_r - \frac{i}{2} \frac{ \text{Re}[Y(\omega_r)]}{C_r}
\end{align}
which implies that: $\kappa_r = \frac{ \text{Re}[Y(\omega_r)]}{C_r}$

This procedure is generally most valid when $\kappa_r \ll \omega_r$, the resonance is not strongly coupled to any nearby modes, and the real part of admittance can be approximated as a constant (meaning that its derivative is small compared to that of the imaginary part: $\text{Re}[Y(\omega_r)]'\ll2C_r$). A general heuristic is that $\kappa_r/2$ should ideally be much smaller than the frequency range over which $(\text{Im}[Y(\omega)])$ is approximately linear, such that this procedure is well-motivated.

We now show more explicitly how the Eq. \eqref{YLinear} corresponds to the near-resonance admittance response of a parallel LCR oscillator (where $\omega_r = 1/\sqrt{LC}$ and $\Delta \omega = \omega - \omega_r$):

\begin{align}
    Y(\omega) &= i \omega C - \frac{i}{ \omega L} +\frac{1}{R} \nonumber \\
            &= \frac{i}{\omega L} \left( \omega^2 LC - 1 \right) +\frac{1}{R}  \nonumber \\
            &= \frac{i C}{\omega} \left( \omega^2  - \omega_r^2 \right) +\frac{1}{R} \nonumber  \\
            & = \frac{i C}{\omega_r+\Delta \omega} \left( (\omega_r+\Delta\omega) ^2  - \omega_r^2 \right) +\frac{1}{R} \nonumber \\
            & \approx 2 i C \left(\Delta\omega \right)+\frac{1}{R}
\end{align}
which has the same form as Eq. \eqref{YLinear} when the resonance frequencies are the same, $C=C_r$, and $\text{Re}[Y(\omega_q)] = 1/R$.

\subsection{Effective capacitance of a CPW $\lambda/2$ resonator} \label{sec:EffectiveCapacitanceCPW}

For distributed-element structures, calculations must be performed to compute the effective capacitance of the synthesized LCR-oscillator model. We examine in particular the case from this work: the $\lambda/2$ mode of an open-terminated coplanar waveguide resonator with length $l_r$, characteristic impedance $Z_0$ and phase velocity $v_p$. If an electromagnetic port is placed from the centerpin to ground at one end of the resonator, the admittance can be expanded around the first resonance frequency $\omega_r = \pi v_p/ l_r$ as \cite{david_m_pozar_microwave_2012}:
\begin{align}
    Y(\omega) &= \frac{i}{Z_0} \tan \left( \frac{\omega l_r}{v_p} \right) \nonumber \\
              &= \frac{i}{Z_0} \tan \left( \frac{\omega l_r}{v_p} -\frac{\omega_r l_r}{v_p} \right) \nonumber \\
              & \approx \frac{i l_r}{Z_0 v_p} \left(\omega - \omega_r \right) \nonumber \\
              & = 2 i\frac{ \pi }{2 \omega_r Z_0 } \left(\omega - \omega_r \right)
\end{align}
Comparing this term with the imaginary part of equation \eqref{YLinear} gives the standard expressions for effective capacitance and inductance from \cite{goppl_coplanar_2008}:
\begin{align}
    C_r &= \frac{\pi}{2 \omega_r Z_0} \\
    \implies L_r &=\frac{1}{C_r \omega_r^2} = \frac{2 Z_0}{\pi \omega_r} 
\end{align}
To bring the admittance expression more fully into the form of \eqref{YLinear}, we add in the external decay that stems from $\text{Re}[Y(\omega_r)]$, computed from the full-wave simulation at the electromagnetic port.

\subsection{Qubit decay rate calculations}

To calculate the decay rate of the transmon qubit, we use the standard weakly-anharmonic approximation, which treats the transmon as an LCR oscillator where the SQUID loop functions as an inductor. In this case, the qubit's decay rate can be calculated as \cite{houck_controlling_2008}:
\begin{align}
\kappa_q \approx \frac{\text{Re}[Y_{q}(\omega_q)]}{C_q}
\end{align}
Here, admittance is measured by placing an electromagnetic port across the SQUID loop. To extract the qubit's capacitance, we take the slope of the imaginary part of admittance, since:
\begin{align}
\text{Im}[Y_{q}(\omega)] \approx i \omega C_q
\end{align}
The Purcell decay limit can then be calculated as $T_1 = 1/\kappa_q$.

\section{Linewidth plateau filters} \label{app:LinewidthPlateauFilters}

\subsection{Resonator decay rate}

We now outline how to calculate the resonator decay rates through a linewidth plateau filter, for the lumped-element model shown in Main Text Fig. \ref{fig:LumpedFilter}. We will highlight the overall rational-polynomial structure of the response, which creates the linewidth plateau.

To calculate the external linewidth through the filter we need to compute the real part of admittance at the filter's input, caused by the dissipation into the output line. For a cascaded structure, this quantity can be calculated using the ABCD matrix \cite{david_m_pozar_microwave_2012}. Each circuit element has its own ABCD matrix, and the overall matrix is calculated by multiplying the matrices together in the order they appear in the circuit. In the end, the matrix will relate the voltage and current at the left ($V_1$ and 
$I_1$) and right ($V_2$ and
$I_2$) sides of the system as:
\begin{align} \label{eq:ABCD}
\begin{bmatrix}
V_1\\
I_1
\end{bmatrix}
=
\begin{bmatrix}
A & B \\
C & D
\end{bmatrix}
\begin{bmatrix}
V_2\\
-I_2
\end{bmatrix}
\end{align}
where all quantities are in the frequency domain. From the entries of this matrix, admittance can subsequently be calculated.

If, as in Fig. \ref{fig:LumpedFilter}, the output impedance $Z_0$ is on the right hand side, then the total ABCD matrix can be written as a multiplication of the filter's ABCD matrix with that of the output impedance:
\begin{align}
\begin{bmatrix}
A & B \\
C & D
\end{bmatrix}
&=
\begin{bmatrix}
A_{f} & B_{f} \\
C_{f} & D_{f}
\end{bmatrix}
\begin{bmatrix}
1 & Z_0 \\
0 & 1
\end{bmatrix} \\
& =
\begin{bmatrix}
A_{f} & A_{f}Z_0 + B_{f} \nonumber \\
C_{f} & C_{f}Z_0 + D_{f}
\end{bmatrix}
\end{align}
Note that since the filter is assumed to be lossless, $A_{f}D_{f} - B_{f}C_{f} = 1$. Using the definition of the ABCD matrix (Eq. \ref{eq:ABCD}), we can then see that when the output port is connected to ground ($V_2 = 0$) the admittance is equal to $Y_r(\omega) = I_1/V_1 = D/B$. Thus we have that:
\begin{align}
Y_r(\omega) &=\frac{D}{B} 
\nonumber \\
            &=\frac{C_{f}Z_0 + D_{f}}{A_{f}Z_0 + B_{f}} \nonumber \\
            & = \frac{ Z_0(A_f D_f - B_f C_f) + (A_f C_f Z_0^2 -B_f D_f)} {A_f^2 Z_0^2 - B_f^2} \nonumber \\
            & = \frac{ Z_0+ (A_f C_f Z_0^2 -B_f D_f)} {A_f^2 Z_0^2 - B_f^2} 
\end{align}

Since $A_{f}$ and $D_{f}$ are real, while $B_{f}$ and $C_{f}$ are imaginary, we have that:

\begin{align} \label{eq:AdmittanceFilter}
\text{Re} [ Y_r(\omega)] &=
            \frac{Z_0}{A_f^2 Z_0^2 -B_f^2} \nonumber \\
            &=
            \frac{Z_0}{|A_f|^2 Z_0^2 +|B_f|^2} 
\end{align}

The ABCD matrix for the filter is generated by multiplying alternating series capacitance and shunt inductance matrices, of the respective forms:
\begin{align}
M_{C_i}
&=
\begin{bmatrix}
    1 & Z_{C_i} \\
    0 & 1
\end{bmatrix} \\
M_{L_i}
&=
\begin{bmatrix}
    1 &  0 \\
    Y_{L_i} & 1
\end{bmatrix}
\end{align}
where $Z_{C_i} = 1/(i \omega C_i)$ and $Y_{L_i} = 1/(i \omega L_i)$.

For a filter whose first coupling segment is capacitive and that has $N$ total capacitive/inductive elements, we then have that $A_f$ and $B_f$ are polynomials in $1/\omega$:
\begin{align}
A_f &= P_{A}(1/\omega) \\
B_f &= i P_{B}(1/\omega)
\end{align}
Here, $A_f$ consists only of even-power terms and $B_f$ has only odd powers, with the overall degrees of the polynomials equal to:
\begin{align}
\text{deg}[P_{A}(1/\omega) ] &= 2 \lfloor N/2 \rfloor \\
\text{deg}[P_{B}(1/\omega) ] &= 2 \lceil N/2 \rceil -1
\end{align}
These properties imply that, after multiplying the numerator and denominator of Eq. \ref{eq:AdmittanceFilter} by $\omega^{2N}/Z_0$, we have:
\begin{align}
\text{Re} [ Y_r(\omega)] &=
            \frac{\omega^{2N}}{P_{N}(\omega^2)} \\
P_{N}(\omega^2) &= (\omega^{2N}/Z_0) (|A_f|^2 Z_0^2 +|B_f|^2)
\end{align}
where $P_{N}(\omega^2)$ is a degree-N polynomial with non-negative value for real $\omega$.

Since the external resonator linewidth is proportional to the real part of admittance, this rational polynomial function largely dictates the resonator's external coupling. At low frequencies it goes as $\omega^{2N}$, and at a high frequencies it approaches a constant value---though this latter feature is unrealistic in practical devices, as the lumped-element approximation will also break down at these high frequencies. By engineering interference between terms of different polynomial orders in the denominator term $P_{N}(\omega^2) $, we can generate different linewidth profiles as a function of resonator frequency to create features such as the linewidth plateau. Zeroes of $P_{N}(\omega^2)$ near the real frequency axis correspond to poles of the system and resonant modes of the filter, which we aim to place above the frequency of the linewidth plateau.

\subsection{Qubit decay rate}

\begin{figure}
    \centering
    \includegraphics[width=1\linewidth]{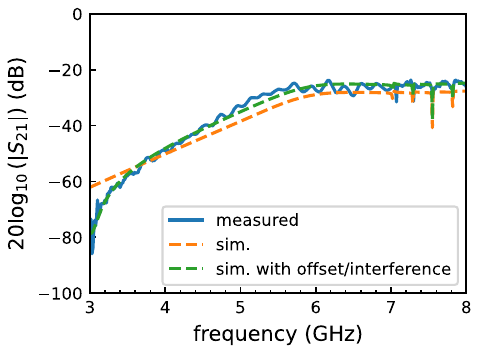}
    \caption{$S_{21}$ measurement and simulation over a broader frequency range. The measured curve is shown in blue alongside the corresponding electromagnetically-simulated result of the layout only (orange). The remaining curve (green) augments the simulated one with additional fit parameters representing offset and interference factors. Interference likely results from parasitic transmission channels in the device or package.}
\label{fig:SupplementalTransition}
\end{figure}

\begin{figure*}
    \centering
    \includegraphics[width=.9\linewidth]{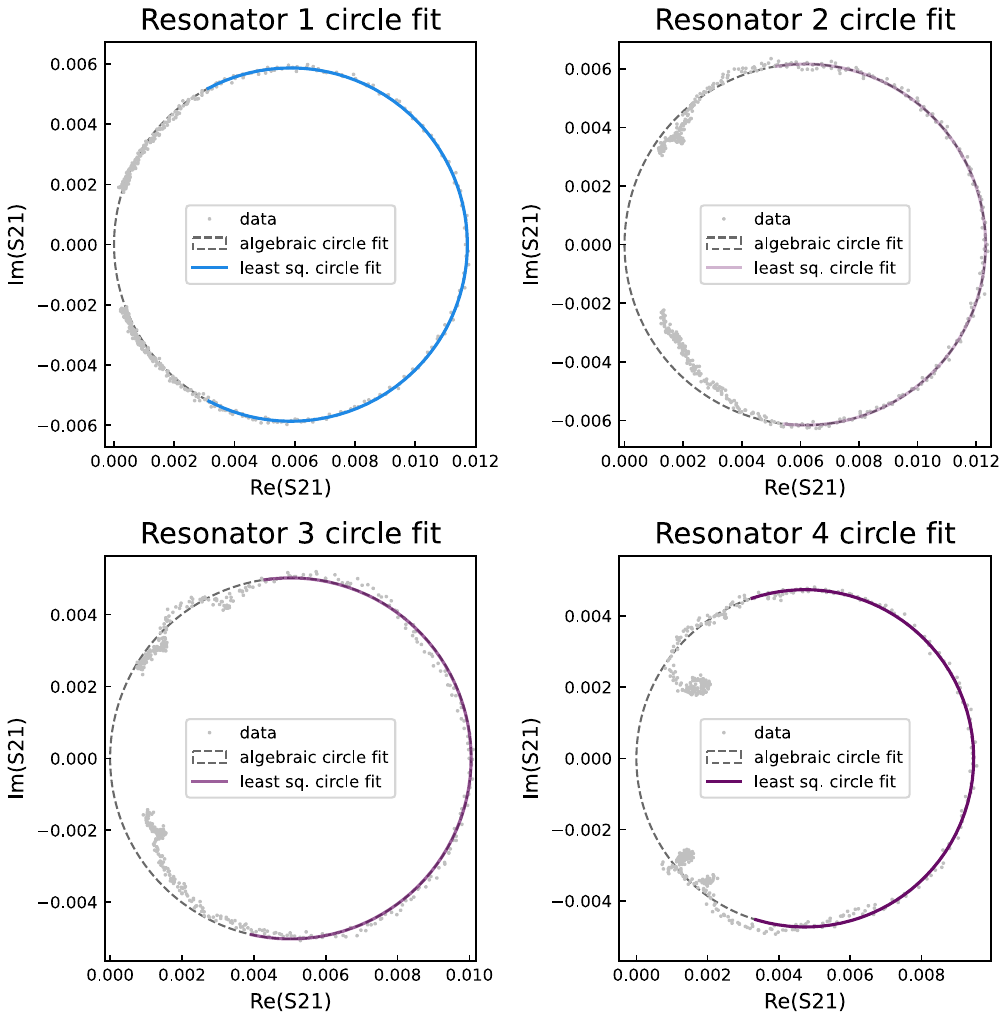}
    \caption{Sample circle fits for each of the four resonators. Translated and rotated data is plotted alongside an algebraic circle fit and a least squares circle fit, both of which are performed on a subset of points around the resonance frequency.} \label{fig:Supplement_circlefits_optimized}
\end{figure*}

As shown in Main Text Fig. \ref{fig:LumpedFilter}, the qubit also experiences external decay through the external readout lines. This decay is filtered both by the qubit's detuning from the readout resonator and by the Purcell filter. We can use a similar expression to calculate the real part of its admittance, using Eq. \ref{eq:AdmittanceFilter} but adding in the filtering effect of the resonator:
\begin{align}
\begin{bmatrix}
A_f' & B_f' \\
C_f' & D_f'
\end{bmatrix} =
\begin{bmatrix}
1 & Z_C \\
0 & 1
\end{bmatrix}
\begin{bmatrix}
1 & 0 \\
Y_r & 1
\end{bmatrix}
\begin{bmatrix}
A_f & B_f \\
C_f & D_f
\end{bmatrix} \nonumber \\
=
\begin{bmatrix}
A_f (1 + Z_C Y_r) + C_f (Z_C) & B_f(1 + Z_C Y_r) + D_f(Z_C) \\
A_f Y_r + C_f & B_f Y_r + D_f
\end{bmatrix}
\end{align}
Here, $Z_C$ is the qubit's coupling capacitance to the readout resonator, while $Y_r$ is the admittance of the readout resonator:
\begin{align}
Z_C &= \frac{1}{i \omega C_c}  \\
Y_r &= i \omega C_r + \frac{1}{i \omega L_r}
\end{align}

Then, as in the external readout decay calculation, we can compute the qubit's real part of admittance as:
\begin{align}
\text{Re} [ Y_q(\omega)] &=
            \frac{\omega^{2N+2}}{P_{N+1}(\omega^2)} \\
P_{N+1}(\omega^2) &= (\omega^{2N+2}/Z_0) (|A_f'|^2 Z_0^2 +|B_f'|^2)
\end{align}
Since the qubit's external decay rate is proportional to this quantity, we thus have that at low frequencies $\kappa_q \propto \omega^{2N+2}$, and so adding additional filter segments can improve qubit protection below the filter's linewidth plateau.

\section{Scattering and circle fitting} \label{sec:ScatteringAndFitting}

\subsection{Scattering profile of the device}

In Fig. \ref{fig:SupplementalTransition} we plot the scattering profile of $S_{21}$ transmission across the device, from the input port to the output port, shown in Main Text Fig. \ref{fig:DeviceLayout}. We plot the measured $S_{21}$ spectrum of the device (blue), with a reference transmission background through the fridge---without the device---subtracted. We then simulate transmission through the device using the AWR AXIEM method of moments solver (orange). Discrepancies between simulation and measurement are likely caused in part by interference through parasitic transmission channels, which can have a large effect when overall transmission is low. To account for these effects, we fit a model (green) whose parameters $\{A,B,C,D \}$ represent an overall offset of the transmission spectrum, as well as interference with a constant-amplitude offset term.
\begin{align}
S_{{21}_\text{tot}}(\omega) = A(S_{21}(\omega)) + B e^{i(C + \omega D)}
\end{align}
The reality of the interference effect is likely more complicated, but this simple model captures much of the observed behavior.

\subsection{Circle fitting} \label{sec:CircleFitting}

To compare experimental results with theoretical predictions, we fit the experimentally measured resonator linewidths from transmission data. After removing the effects of cable delay, the transmission spectrum of a resonator should approximate a circle in the complex plane, whose properties encode the resonator's parameters. For a given $S_{21}$ trace, we first perform an algebraic circle fit \cite{taubin_estimation_1991}, and then carry out a translation/rotation of the data such that the circle's center now lies at $(R,0)$, and the point at resonance near $(2R,0)$. This translation then simplifies the process of performing a least-squares circle fit to the data, using the fitting parameters listed in the function below (a re-expression of the standard formula given in \cite{probst_efficient_2015}):
\begin{align}
    S_{21}(\omega) = A_\text{r} + i A_\text{i} + \frac{2R e^{i \phi}}{1+\frac{i}{\pi \kappa}(\omega-\omega_r)}
\end{align}
where $\omega = 2 \pi f$. Note that due to the translation/rotation, the correction factors $A_\text{r}$, $A_\text{i}$, and $\phi$ should be small. From this fit we (most importantly) extract the resonator linewidth $\kappa$ and the (angular) resonance frequency $\omega_r$. 

Examples of this procedure are shown in Fig. \ref{fig:Supplement_circlefits_optimized}, with the data, algebraic circle fit, and least squares circle fits (over a limited frequency range) plotted. Due to impedance mismatches in the package and between the package and chip, significant deviations from the ideal circular form are present in resonators $2$, $3$, and $4$. Given this systematic uncertainty, we performed least squares circle fits for a range of frequency windows around resonance, reporting a weighted average $\kappa$ and standard deviation in Table \ref{table:ResFreq}.

\bibliography{references.bib}

\begin{thebibliography}{28}%
\makeatletter
\providecommand \@ifxundefined [1]{%
 \@ifx{#1\undefined}
}%
\providecommand \@ifnum [1]{%
 \ifnum #1\expandafter \@firstoftwo
 \else \expandafter \@secondoftwo
 \fi
}%
\providecommand \@ifx [1]{%
 \ifx #1\expandafter \@firstoftwo
 \else \expandafter \@secondoftwo
 \fi
}%
\providecommand \natexlab [1]{#1}%
\providecommand \enquote  [1]{``#1''}%
\providecommand \bibnamefont  [1]{#1}%
\providecommand \bibfnamefont [1]{#1}%
\providecommand \citenamefont [1]{#1}%
\providecommand \href@noop [0]{\@secondoftwo}%
\providecommand \href [0]{\begingroup \@sanitize@url \@href}%
\providecommand \@href[1]{\@@startlink{#1}\@@href}%
\providecommand \@@href[1]{\endgroup#1\@@endlink}%
\providecommand \@sanitize@url [0]{\catcode `\\12\catcode `\$12\catcode
  `\&12\catcode `\#12\catcode `\^12\catcode `\_12\catcode `\%12\relax}%
\providecommand \@@startlink[1]{}%
\providecommand \@@endlink[0]{}%
\providecommand \url  [0]{\begingroup\@sanitize@url \@url }%
\providecommand \@url [1]{\endgroup\@href {#1}{\urlprefix }}%
\providecommand \urlprefix  [0]{URL }%
\providecommand \Eprint [0]{\href }%
\providecommand \doibase [0]{https://doi.org/}%
\providecommand \selectlanguage [0]{\@gobble}%
\providecommand \bibinfo  [0]{\@secondoftwo}%
\providecommand \bibfield  [0]{\@secondoftwo}%
\providecommand \translation [1]{[#1]}%
\providecommand \BibitemOpen [0]{}%
\providecommand \bibitemStop [0]{}%
\providecommand \bibitemNoStop [0]{.\EOS\space}%
\providecommand \EOS [0]{\spacefactor3000\relax}%
\providecommand \BibitemShut  [1]{\csname bibitem#1\endcsname}%
\let\auto@bib@innerbib\@empty
\bibitem [{\citenamefont {{Purcell, E.
  M.}}(1946)}]{purcell_e_m_proceedings_1946}%
  \BibitemOpen
  \bibfield  {author} {\bibinfo {author} {\bibnamefont {{Purcell, E. M.}}},\
  }\bibfield  {title} {\bibinfo {title} {Proceedings of the {American}
  {Physical} {Society}},\ }\href {https://doi.org/10.1103/PhysRev.69.674.2}
  {\bibfield  {journal} {\bibinfo  {journal} {Physical Review}\ }\textbf
  {\bibinfo {volume} {69}},\ \bibinfo {pages} {674} (\bibinfo {year}
  {1946})}\BibitemShut {NoStop}%
\bibitem [{\citenamefont {{David M.
  Pozar}}(2012)}]{david_m_pozar_microwave_2012}%
  \BibitemOpen
  \bibfield  {author} {\bibinfo {author} {\bibnamefont {{David M. Pozar}}},\
  }\href@noop {} {\emph {\bibinfo {title} {Microwave {Engineering}}}},\
  \bibinfo {edition} {4th}\ ed.\ (\bibinfo  {publisher} {John Wiley and Sons},\
  \bibinfo {year} {2012})\BibitemShut {NoStop}%
\bibitem [{\citenamefont {Reed}\ \emph {et~al.}(2010)\citenamefont {Reed},
  \citenamefont {Johnson}, \citenamefont {Houck}, \citenamefont {DiCarlo},
  \citenamefont {Chow}, \citenamefont {Schuster}, \citenamefont {Frunzio},\
  and\ \citenamefont {Schoelkopf}}]{reed_fast_2010}%
  \BibitemOpen
  \bibfield  {author} {\bibinfo {author} {\bibfnamefont {M.~D.}\ \bibnamefont
  {Reed}}, \bibinfo {author} {\bibfnamefont {B.~R.}\ \bibnamefont {Johnson}},
  \bibinfo {author} {\bibfnamefont {A.~A.}\ \bibnamefont {Houck}}, \bibinfo
  {author} {\bibfnamefont {L.}~\bibnamefont {DiCarlo}}, \bibinfo {author}
  {\bibfnamefont {J.~M.}\ \bibnamefont {Chow}}, \bibinfo {author}
  {\bibfnamefont {D.~I.}\ \bibnamefont {Schuster}}, \bibinfo {author}
  {\bibfnamefont {L.}~\bibnamefont {Frunzio}},\ and\ \bibinfo {author}
  {\bibfnamefont {R.~J.}\ \bibnamefont {Schoelkopf}},\ }\bibfield  {title}
  {\bibinfo {title} {Fast reset and suppressing spontaneous emission of a
  superconducting qubit},\ }\href {https://doi.org/10.1063/1.3435463}
  {\bibfield  {journal} {\bibinfo  {journal} {Applied Physics Letters}\
  }\textbf {\bibinfo {volume} {96}},\ \bibinfo {pages} {203110} (\bibinfo
  {year} {2010})}\BibitemShut {NoStop}%
\bibitem [{\citenamefont {Sete}\ \emph {et~al.}(2015)\citenamefont {Sete},
  \citenamefont {Martinis},\ and\ \citenamefont
  {Korotkov}}]{sete_quantum_2015}%
  \BibitemOpen
  \bibfield  {author} {\bibinfo {author} {\bibfnamefont {E.~A.}\ \bibnamefont
  {Sete}}, \bibinfo {author} {\bibfnamefont {J.~M.}\ \bibnamefont {Martinis}},\
  and\ \bibinfo {author} {\bibfnamefont {A.~N.}\ \bibnamefont {Korotkov}},\
  }\bibfield  {title} {\bibinfo {title} {Quantum theory of a bandpass {Purcell}
  filter for qubit readout},\ }\href
  {https://doi.org/10.1103/PhysRevA.92.012325} {\bibfield  {journal} {\bibinfo
  {journal} {Physical Review A}\ }\textbf {\bibinfo {volume} {92}},\ \bibinfo
  {pages} {012325} (\bibinfo {year} {2015})}\BibitemShut {NoStop}%
\bibitem [{\citenamefont {Jeffrey}\ \emph {et~al.}(2014)\citenamefont
  {Jeffrey}, \citenamefont {Sank}, \citenamefont {Mutus}, \citenamefont
  {White}, \citenamefont {Kelly}, \citenamefont {Barends}, \citenamefont
  {Chen}, \citenamefont {Chen}, \citenamefont {Chiaro}, \citenamefont
  {Dunsworth}, \citenamefont {Megrant}, \citenamefont {O’Malley},
  \citenamefont {Neill}, \citenamefont {Roushan}, \citenamefont {Vainsencher},
  \citenamefont {Wenner}, \citenamefont {Cleland},\ and\ \citenamefont
  {Martinis}}]{jeffrey_fast_2014}%
  \BibitemOpen
  \bibfield  {author} {\bibinfo {author} {\bibfnamefont {E.}~\bibnamefont
  {Jeffrey}}, \bibinfo {author} {\bibfnamefont {D.}~\bibnamefont {Sank}},
  \bibinfo {author} {\bibfnamefont {J.}~\bibnamefont {Mutus}}, \bibinfo
  {author} {\bibfnamefont {T.}~\bibnamefont {White}}, \bibinfo {author}
  {\bibfnamefont {J.}~\bibnamefont {Kelly}}, \bibinfo {author} {\bibfnamefont
  {R.}~\bibnamefont {Barends}}, \bibinfo {author} {\bibfnamefont
  {Y.}~\bibnamefont {Chen}}, \bibinfo {author} {\bibfnamefont {Z.}~\bibnamefont
  {Chen}}, \bibinfo {author} {\bibfnamefont {B.}~\bibnamefont {Chiaro}},
  \bibinfo {author} {\bibfnamefont {A.}~\bibnamefont {Dunsworth}}, \bibinfo
  {author} {\bibfnamefont {A.}~\bibnamefont {Megrant}}, \bibinfo {author}
  {\bibfnamefont {P.}~\bibnamefont {O’Malley}}, \bibinfo {author}
  {\bibfnamefont {C.}~\bibnamefont {Neill}}, \bibinfo {author} {\bibfnamefont
  {P.}~\bibnamefont {Roushan}}, \bibinfo {author} {\bibfnamefont
  {A.}~\bibnamefont {Vainsencher}}, \bibinfo {author} {\bibfnamefont
  {J.}~\bibnamefont {Wenner}}, \bibinfo {author} {\bibfnamefont
  {A.}~\bibnamefont {Cleland}},\ and\ \bibinfo {author} {\bibfnamefont {J.~M.}\
  \bibnamefont {Martinis}},\ }\bibfield  {title} {\bibinfo {title} {Fast
  {Accurate} {State} {Measurement} with {Superconducting} {Qubits}},\ }\href
  {https://doi.org/10.1103/PhysRevLett.112.190504} {\bibfield  {journal}
  {\bibinfo  {journal} {Physical Review Letters}\ }\textbf {\bibinfo {volume}
  {112}},\ \bibinfo {pages} {190504} (\bibinfo {year} {2014})}\BibitemShut
  {NoStop}%
\bibitem [{\citenamefont {Walter}\ \emph {et~al.}(2017)\citenamefont {Walter},
  \citenamefont {Kurpiers}, \citenamefont {Gasparinetti}, \citenamefont
  {Magnard}, \citenamefont {Potočnik}, \citenamefont {Salathé}, \citenamefont
  {Pechal}, \citenamefont {Mondal}, \citenamefont {Oppliger}, \citenamefont
  {Eichler},\ and\ \citenamefont {Wallraff}}]{walter_rapid_2017}%
  \BibitemOpen
  \bibfield  {author} {\bibinfo {author} {\bibfnamefont {T.}~\bibnamefont
  {Walter}}, \bibinfo {author} {\bibfnamefont {P.}~\bibnamefont {Kurpiers}},
  \bibinfo {author} {\bibfnamefont {S.}~\bibnamefont {Gasparinetti}}, \bibinfo
  {author} {\bibfnamefont {P.}~\bibnamefont {Magnard}}, \bibinfo {author}
  {\bibfnamefont {A.}~\bibnamefont {Potočnik}}, \bibinfo {author}
  {\bibfnamefont {Y.}~\bibnamefont {Salathé}}, \bibinfo {author}
  {\bibfnamefont {M.}~\bibnamefont {Pechal}}, \bibinfo {author} {\bibfnamefont
  {M.}~\bibnamefont {Mondal}}, \bibinfo {author} {\bibfnamefont
  {M.}~\bibnamefont {Oppliger}}, \bibinfo {author} {\bibfnamefont
  {C.}~\bibnamefont {Eichler}},\ and\ \bibinfo {author} {\bibfnamefont
  {A.}~\bibnamefont {Wallraff}},\ }\bibfield  {title} {\bibinfo {title} {Rapid
  {High}-{Fidelity} {Single}-{Shot} {Dispersive} {Readout} of {Superconducting}
  {Qubits}},\ }\href {https://doi.org/10.1103/PhysRevApplied.7.054020}
  {\bibfield  {journal} {\bibinfo  {journal} {Physical Review Applied}\
  }\textbf {\bibinfo {volume} {7}},\ \bibinfo {pages} {054020} (\bibinfo {year}
  {2017})}\BibitemShut {NoStop}%
\bibitem [{\citenamefont {Heinsoo}\ \emph {et~al.}(2018)\citenamefont
  {Heinsoo}, \citenamefont {Andersen}, \citenamefont {Remm}, \citenamefont
  {Krinner}, \citenamefont {Walter}, \citenamefont {Salathé}, \citenamefont
  {Gasparinetti}, \citenamefont {Besse}, \citenamefont {Potočnik},
  \citenamefont {Wallraff},\ and\ \citenamefont
  {Eichler}}]{heinsoo_rapid_2018}%
  \BibitemOpen
  \bibfield  {author} {\bibinfo {author} {\bibfnamefont {J.}~\bibnamefont
  {Heinsoo}}, \bibinfo {author} {\bibfnamefont {C.~K.}\ \bibnamefont
  {Andersen}}, \bibinfo {author} {\bibfnamefont {A.}~\bibnamefont {Remm}},
  \bibinfo {author} {\bibfnamefont {S.}~\bibnamefont {Krinner}}, \bibinfo
  {author} {\bibfnamefont {T.}~\bibnamefont {Walter}}, \bibinfo {author}
  {\bibfnamefont {Y.}~\bibnamefont {Salathé}}, \bibinfo {author}
  {\bibfnamefont {S.}~\bibnamefont {Gasparinetti}}, \bibinfo {author}
  {\bibfnamefont {J.-C.}\ \bibnamefont {Besse}}, \bibinfo {author}
  {\bibfnamefont {A.}~\bibnamefont {Potočnik}}, \bibinfo {author}
  {\bibfnamefont {A.}~\bibnamefont {Wallraff}},\ and\ \bibinfo {author}
  {\bibfnamefont {C.}~\bibnamefont {Eichler}},\ }\bibfield  {title} {\bibinfo
  {title} {Rapid {High}-fidelity {Multiplexed} {Readout} of {Superconducting}
  {Qubits}},\ }\href {https://doi.org/10.1103/PhysRevApplied.10.034040}
  {\bibfield  {journal} {\bibinfo  {journal} {Physical Review Applied}\
  }\textbf {\bibinfo {volume} {10}},\ \bibinfo {pages} {034040} (\bibinfo
  {year} {2018})}\BibitemShut {NoStop}%
\bibitem [{\citenamefont {Bronn}\ \emph {et~al.}(2015)\citenamefont {Bronn},
  \citenamefont {Liu}, \citenamefont {Hertzberg}, \citenamefont {Córcoles},
  \citenamefont {Houck}, \citenamefont {Gambetta},\ and\ \citenamefont
  {Chow}}]{bronn_broadband_2015}%
  \BibitemOpen
  \bibfield  {author} {\bibinfo {author} {\bibfnamefont {N.~T.}\ \bibnamefont
  {Bronn}}, \bibinfo {author} {\bibfnamefont {Y.}~\bibnamefont {Liu}}, \bibinfo
  {author} {\bibfnamefont {J.~B.}\ \bibnamefont {Hertzberg}}, \bibinfo {author}
  {\bibfnamefont {A.~D.}\ \bibnamefont {Córcoles}}, \bibinfo {author}
  {\bibfnamefont {A.~A.}\ \bibnamefont {Houck}}, \bibinfo {author}
  {\bibfnamefont {J.~M.}\ \bibnamefont {Gambetta}},\ and\ \bibinfo {author}
  {\bibfnamefont {J.~M.}\ \bibnamefont {Chow}},\ }\bibfield  {title} {\bibinfo
  {title} {Broadband filters for abatement of spontaneous emission in circuit
  quantum electrodynamics},\ }\href {https://doi.org/10.1063/1.4934867}
  {\bibfield  {journal} {\bibinfo  {journal} {Applied Physics Letters}\
  }\textbf {\bibinfo {volume} {107}},\ \bibinfo {pages} {172601} (\bibinfo
  {year} {2015})}\BibitemShut {NoStop}%
\bibitem [{\citenamefont {Cleland}\ \emph {et~al.}(2019)\citenamefont
  {Cleland}, \citenamefont {Pechal}, \citenamefont {Stas}, \citenamefont
  {Sarabalis}, \citenamefont {Wollack},\ and\ \citenamefont
  {Safavi-Naeini}}]{cleland_mechanical_2019}%
  \BibitemOpen
  \bibfield  {author} {\bibinfo {author} {\bibfnamefont {A.~Y.}\ \bibnamefont
  {Cleland}}, \bibinfo {author} {\bibfnamefont {M.}~\bibnamefont {Pechal}},
  \bibinfo {author} {\bibfnamefont {P.-J.~C.}\ \bibnamefont {Stas}}, \bibinfo
  {author} {\bibfnamefont {C.~J.}\ \bibnamefont {Sarabalis}}, \bibinfo {author}
  {\bibfnamefont {E.~A.}\ \bibnamefont {Wollack}},\ and\ \bibinfo {author}
  {\bibfnamefont {A.~H.}\ \bibnamefont {Safavi-Naeini}},\ }\bibfield  {title}
  {\bibinfo {title} {Mechanical {Purcell} filters for microwave quantum
  machines},\ }\href {https://doi.org/10.1063/1.5111151} {\bibfield  {journal}
  {\bibinfo  {journal} {Applied Physics Letters}\ }\textbf {\bibinfo {volume}
  {115}},\ \bibinfo {pages} {263504} (\bibinfo {year} {2019})}\BibitemShut
  {NoStop}%
\bibitem [{\citenamefont {Yan}\ \emph {et~al.}(2023)\citenamefont {Yan},
  \citenamefont {Wu}, \citenamefont {Lingenfelter}, \citenamefont {Joshi},
  \citenamefont {Andersson}, \citenamefont {Conner}, \citenamefont {Chou},
  \citenamefont {Grebel}, \citenamefont {Miller}, \citenamefont {Povey},
  \citenamefont {Qiao}, \citenamefont {Clerk},\ and\ \citenamefont
  {Cleland}}]{yan_broadband_2023}%
  \BibitemOpen
  \bibfield  {author} {\bibinfo {author} {\bibfnamefont {H.}~\bibnamefont
  {Yan}}, \bibinfo {author} {\bibfnamefont {X.}~\bibnamefont {Wu}}, \bibinfo
  {author} {\bibfnamefont {A.}~\bibnamefont {Lingenfelter}}, \bibinfo {author}
  {\bibfnamefont {Y.~J.}\ \bibnamefont {Joshi}}, \bibinfo {author}
  {\bibfnamefont {G.}~\bibnamefont {Andersson}}, \bibinfo {author}
  {\bibfnamefont {C.~R.}\ \bibnamefont {Conner}}, \bibinfo {author}
  {\bibfnamefont {M.-H.}\ \bibnamefont {Chou}}, \bibinfo {author}
  {\bibfnamefont {J.}~\bibnamefont {Grebel}}, \bibinfo {author} {\bibfnamefont
  {J.~M.}\ \bibnamefont {Miller}}, \bibinfo {author} {\bibfnamefont {R.~G.}\
  \bibnamefont {Povey}}, \bibinfo {author} {\bibfnamefont {H.}~\bibnamefont
  {Qiao}}, \bibinfo {author} {\bibfnamefont {A.~A.}\ \bibnamefont {Clerk}},\
  and\ \bibinfo {author} {\bibfnamefont {A.~N.}\ \bibnamefont {Cleland}},\
  }\bibfield  {title} {\bibinfo {title} {Broadband bandpass {Purcell} filter
  for circuit quantum electrodynamics},\ }\href
  {https://doi.org/10.1063/5.0161893} {\bibfield  {journal} {\bibinfo
  {journal} {Applied Physics Letters}\ }\textbf {\bibinfo {volume} {123}},\
  \bibinfo {pages} {134001} (\bibinfo {year} {2023})}\BibitemShut {NoStop}%
\bibitem [{\citenamefont {Park}\ \emph {et~al.}(2024)\citenamefont {Park},
  \citenamefont {Choi}, \citenamefont {Kim}, \citenamefont {Jo}, \citenamefont
  {Lee}, \citenamefont {Kim}, \citenamefont {Park}, \citenamefont {Lee},\ and\
  \citenamefont {Hahn}}]{park_characterization_2024}%
  \BibitemOpen
  \bibfield  {author} {\bibinfo {author} {\bibfnamefont {S.~H.}\ \bibnamefont
  {Park}}, \bibinfo {author} {\bibfnamefont {G.}~\bibnamefont {Choi}}, \bibinfo
  {author} {\bibfnamefont {G.}~\bibnamefont {Kim}}, \bibinfo {author}
  {\bibfnamefont {J.}~\bibnamefont {Jo}}, \bibinfo {author} {\bibfnamefont
  {B.}~\bibnamefont {Lee}}, \bibinfo {author} {\bibfnamefont {G.}~\bibnamefont
  {Kim}}, \bibinfo {author} {\bibfnamefont {K.}~\bibnamefont {Park}}, \bibinfo
  {author} {\bibfnamefont {Y.-H.}\ \bibnamefont {Lee}},\ and\ \bibinfo {author}
  {\bibfnamefont {S.}~\bibnamefont {Hahn}},\ }\bibfield  {title} {\bibinfo
  {title} {Characterization of broadband {Purcell} filters with compact
  footprint for fast multiplexed superconducting qubit readout},\ }\href
  {https://doi.org/10.1063/5.0182642} {\bibfield  {journal} {\bibinfo
  {journal} {Applied Physics Letters}\ }\textbf {\bibinfo {volume} {124}},\
  \bibinfo {pages} {044003} (\bibinfo {year} {2024})}\BibitemShut {NoStop}%
\bibitem [{\citenamefont {Narla}\ \emph {et~al.}(2016)\citenamefont {Narla},
  \citenamefont {Shankar}, \citenamefont {Hatridge}, \citenamefont {Leghtas},
  \citenamefont {Sliwa}, \citenamefont {Zalys-Geller}, \citenamefont
  {Mundhada}, \citenamefont {Pfaff}, \citenamefont {Frunzio}, \citenamefont
  {Schoelkopf},\ and\ \citenamefont {Devoret}}]{narla_robust_2016}%
  \BibitemOpen
  \bibfield  {author} {\bibinfo {author} {\bibfnamefont {A.}~\bibnamefont
  {Narla}}, \bibinfo {author} {\bibfnamefont {S.}~\bibnamefont {Shankar}},
  \bibinfo {author} {\bibfnamefont {M.}~\bibnamefont {Hatridge}}, \bibinfo
  {author} {\bibfnamefont {Z.}~\bibnamefont {Leghtas}}, \bibinfo {author}
  {\bibfnamefont {K.}~\bibnamefont {Sliwa}}, \bibinfo {author} {\bibfnamefont
  {E.}~\bibnamefont {Zalys-Geller}}, \bibinfo {author} {\bibfnamefont
  {S.}~\bibnamefont {Mundhada}}, \bibinfo {author} {\bibfnamefont
  {W.}~\bibnamefont {Pfaff}}, \bibinfo {author} {\bibfnamefont
  {L.}~\bibnamefont {Frunzio}}, \bibinfo {author} {\bibfnamefont
  {R.}~\bibnamefont {Schoelkopf}},\ and\ \bibinfo {author} {\bibfnamefont
  {M.}~\bibnamefont {Devoret}},\ }\bibfield  {title} {\bibinfo {title} {Robust
  {Concurrent} {Remote} {Entanglement} {Between} {Two} {Superconducting}
  {Qubits}},\ }\href {https://doi.org/10.1103/PhysRevX.6.031036} {\bibfield
  {journal} {\bibinfo  {journal} {Physical Review X}\ }\textbf {\bibinfo
  {volume} {6}},\ \bibinfo {pages} {031036} (\bibinfo {year}
  {2016})}\BibitemShut {NoStop}%
\bibitem [{\citenamefont {Wang}\ \emph {et~al.}(2019)\citenamefont {Wang},
  \citenamefont {Shankar}, \citenamefont {Minev}, \citenamefont
  {Campagne-Ibarcq}, \citenamefont {Narla},\ and\ \citenamefont
  {Devoret}}]{wang_cavity_2019}%
  \BibitemOpen
  \bibfield  {author} {\bibinfo {author} {\bibfnamefont {Z.}~\bibnamefont
  {Wang}}, \bibinfo {author} {\bibfnamefont {S.}~\bibnamefont {Shankar}},
  \bibinfo {author} {\bibfnamefont {Z.}~\bibnamefont {Minev}}, \bibinfo
  {author} {\bibfnamefont {P.}~\bibnamefont {Campagne-Ibarcq}}, \bibinfo
  {author} {\bibfnamefont {A.}~\bibnamefont {Narla}},\ and\ \bibinfo {author}
  {\bibfnamefont {M.}~\bibnamefont {Devoret}},\ }\bibfield  {title} {\bibinfo
  {title} {Cavity {Attenuators} for {Superconducting} {Qubits}},\ }\href
  {https://doi.org/10.1103/PhysRevApplied.11.014031} {\bibfield  {journal}
  {\bibinfo  {journal} {Physical Review Applied}\ }\textbf {\bibinfo {volume}
  {11}},\ \bibinfo {pages} {014031} (\bibinfo {year} {2019})}\BibitemShut
  {NoStop}%
\bibitem [{\citenamefont {Houck}\ \emph {et~al.}(2008)\citenamefont {Houck},
  \citenamefont {Schreier}, \citenamefont {Johnson}, \citenamefont {Chow},
  \citenamefont {Koch}, \citenamefont {Gambetta}, \citenamefont {Schuster},
  \citenamefont {Frunzio}, \citenamefont {Devoret}, \citenamefont {Girvin},\
  and\ \citenamefont {Schoelkopf}}]{houck_controlling_2008}%
  \BibitemOpen
  \bibfield  {author} {\bibinfo {author} {\bibfnamefont {A.~A.}\ \bibnamefont
  {Houck}}, \bibinfo {author} {\bibfnamefont {J.~A.}\ \bibnamefont {Schreier}},
  \bibinfo {author} {\bibfnamefont {B.~R.}\ \bibnamefont {Johnson}}, \bibinfo
  {author} {\bibfnamefont {J.~M.}\ \bibnamefont {Chow}}, \bibinfo {author}
  {\bibfnamefont {J.}~\bibnamefont {Koch}}, \bibinfo {author} {\bibfnamefont
  {J.~M.}\ \bibnamefont {Gambetta}}, \bibinfo {author} {\bibfnamefont {D.~I.}\
  \bibnamefont {Schuster}}, \bibinfo {author} {\bibfnamefont {L.}~\bibnamefont
  {Frunzio}}, \bibinfo {author} {\bibfnamefont {M.~H.}\ \bibnamefont
  {Devoret}}, \bibinfo {author} {\bibfnamefont {S.~M.}\ \bibnamefont
  {Girvin}},\ and\ \bibinfo {author} {\bibfnamefont {R.~J.}\ \bibnamefont
  {Schoelkopf}},\ }\bibfield  {title} {\bibinfo {title} {Controlling the
  {Spontaneous} {Emission} of a {Superconducting} {Transmon} {Qubit}},\ }\href
  {https://doi.org/10.1103/PhysRevLett.101.080502} {\bibfield  {journal}
  {\bibinfo  {journal} {Physical Review Letters}\ }\textbf {\bibinfo {volume}
  {101}},\ \bibinfo {pages} {080502} (\bibinfo {year} {2008})}\BibitemShut
  {NoStop}%
\bibitem [{\citenamefont {Sunada}\ \emph {et~al.}(2022)\citenamefont {Sunada},
  \citenamefont {Kono}, \citenamefont {Ilves}, \citenamefont {Tamate},
  \citenamefont {Sugiyama}, \citenamefont {Tabuchi},\ and\ \citenamefont
  {Nakamura}}]{sunada_fast_2022}%
  \BibitemOpen
  \bibfield  {author} {\bibinfo {author} {\bibfnamefont {Y.}~\bibnamefont
  {Sunada}}, \bibinfo {author} {\bibfnamefont {S.}~\bibnamefont {Kono}},
  \bibinfo {author} {\bibfnamefont {J.}~\bibnamefont {Ilves}}, \bibinfo
  {author} {\bibfnamefont {S.}~\bibnamefont {Tamate}}, \bibinfo {author}
  {\bibfnamefont {T.}~\bibnamefont {Sugiyama}}, \bibinfo {author}
  {\bibfnamefont {Y.}~\bibnamefont {Tabuchi}},\ and\ \bibinfo {author}
  {\bibfnamefont {Y.}~\bibnamefont {Nakamura}},\ }\bibfield  {title} {\bibinfo
  {title} {Fast {Readout} and {Reset} of a {Superconducting} {Qubit} {Coupled}
  to a {Resonator} with an {Intrinsic} {Purcell} {Filter}},\ }\href
  {https://doi.org/10.1103/PhysRevApplied.17.044016} {\bibfield  {journal}
  {\bibinfo  {journal} {Physical Review Applied}\ }\textbf {\bibinfo {volume}
  {17}},\ \bibinfo {pages} {044016} (\bibinfo {year} {2022})}\BibitemShut
  {NoStop}%
\bibitem [{\citenamefont {Spring}\ \emph {et~al.}(2024)\citenamefont {Spring},
  \citenamefont {Milanovic}, \citenamefont {Sunada}, \citenamefont {Wang},
  \citenamefont {Loo}, \citenamefont {Tamate},\ and\ \citenamefont
  {Nakamura}}]{spring_fast_2024}%
  \BibitemOpen
  \bibfield  {author} {\bibinfo {author} {\bibfnamefont {P.~A.}\ \bibnamefont
  {Spring}}, \bibinfo {author} {\bibfnamefont {L.}~\bibnamefont {Milanovic}},
  \bibinfo {author} {\bibfnamefont {Y.}~\bibnamefont {Sunada}}, \bibinfo
  {author} {\bibfnamefont {S.}~\bibnamefont {Wang}}, \bibinfo {author}
  {\bibfnamefont {A.~F.~v.}\ \bibnamefont {Loo}}, \bibinfo {author}
  {\bibfnamefont {S.}~\bibnamefont {Tamate}},\ and\ \bibinfo {author}
  {\bibfnamefont {Y.}~\bibnamefont {Nakamura}},\ }\href
  {https://doi.org/10.48550/arXiv.2409.04967} {\bibinfo {title} {Fast
  multiplexed superconducting qubit readout with intrinsic {Purcell}
  filtering}} (\bibinfo {year} {2024}),\ \bibinfo {note} {arXiv:2409.04967
  [quant-ph]}\BibitemShut {NoStop}%
\bibitem [{\citenamefont {Yen}\ \emph {et~al.}(2024)\citenamefont {Yen},
  \citenamefont {Ye}, \citenamefont {Peng}, \citenamefont {Wang}, \citenamefont
  {Cunningham}, \citenamefont {Gingras}, \citenamefont {Niedzielski},
  \citenamefont {Stickler}, \citenamefont {Serniak}, \citenamefont {Schwartz},\
  and\ \citenamefont {O'Brien}}]{yen_interferometric_2024}%
  \BibitemOpen
  \bibfield  {author} {\bibinfo {author} {\bibfnamefont {A.}~\bibnamefont
  {Yen}}, \bibinfo {author} {\bibfnamefont {Y.}~\bibnamefont {Ye}}, \bibinfo
  {author} {\bibfnamefont {K.}~\bibnamefont {Peng}}, \bibinfo {author}
  {\bibfnamefont {J.}~\bibnamefont {Wang}}, \bibinfo {author} {\bibfnamefont
  {G.}~\bibnamefont {Cunningham}}, \bibinfo {author} {\bibfnamefont
  {M.}~\bibnamefont {Gingras}}, \bibinfo {author} {\bibfnamefont {B.~M.}\
  \bibnamefont {Niedzielski}}, \bibinfo {author} {\bibfnamefont
  {H.}~\bibnamefont {Stickler}}, \bibinfo {author} {\bibfnamefont
  {K.}~\bibnamefont {Serniak}}, \bibinfo {author} {\bibfnamefont {M.~E.}\
  \bibnamefont {Schwartz}},\ and\ \bibinfo {author} {\bibfnamefont {K.~P.}\
  \bibnamefont {O'Brien}},\ }\href {https://doi.org/10.48550/arXiv.2405.10107}
  {\bibinfo {title} {Interferometric {Purcell} suppression of spontaneous
  emission in a superconducting qubit}} (\bibinfo {year} {2024}),\ \bibinfo
  {note} {arXiv:2405.10107}\BibitemShut {NoStop}%
\bibitem [{\citenamefont {Göppl}\ \emph {et~al.}(2008)\citenamefont {Göppl},
  \citenamefont {Fragner}, \citenamefont {Baur}, \citenamefont {Bianchetti},
  \citenamefont {Filipp}, \citenamefont {Fink}, \citenamefont {Leek},
  \citenamefont {Puebla}, \citenamefont {Steffen},\ and\ \citenamefont
  {Wallraff}}]{goppl_coplanar_2008}%
  \BibitemOpen
  \bibfield  {author} {\bibinfo {author} {\bibfnamefont {M.}~\bibnamefont
  {Göppl}}, \bibinfo {author} {\bibfnamefont {A.}~\bibnamefont {Fragner}},
  \bibinfo {author} {\bibfnamefont {M.}~\bibnamefont {Baur}}, \bibinfo {author}
  {\bibfnamefont {R.}~\bibnamefont {Bianchetti}}, \bibinfo {author}
  {\bibfnamefont {S.}~\bibnamefont {Filipp}}, \bibinfo {author} {\bibfnamefont
  {J.~M.}\ \bibnamefont {Fink}}, \bibinfo {author} {\bibfnamefont {P.~J.}\
  \bibnamefont {Leek}}, \bibinfo {author} {\bibfnamefont {G.}~\bibnamefont
  {Puebla}}, \bibinfo {author} {\bibfnamefont {L.}~\bibnamefont {Steffen}},\
  and\ \bibinfo {author} {\bibfnamefont {A.}~\bibnamefont {Wallraff}},\
  }\bibfield  {title} {\bibinfo {title} {Coplanar waveguide resonators for
  circuit quantum electrodynamics},\ }\href {https://doi.org/10.1063/1.3010859}
  {\bibfield  {journal} {\bibinfo  {journal} {Journal of Applied Physics}\
  }\textbf {\bibinfo {volume} {104}},\ \bibinfo {pages} {113904} (\bibinfo
  {year} {2008})}\BibitemShut {NoStop}%
\bibitem [{\citenamefont {Wisbey}\ \emph {et~al.}(2014)\citenamefont {Wisbey},
  \citenamefont {Martin}, \citenamefont {Reinisch},\ and\ \citenamefont
  {Gao}}]{wisbey_new_2014}%
  \BibitemOpen
  \bibfield  {author} {\bibinfo {author} {\bibfnamefont {D.~S.}\ \bibnamefont
  {Wisbey}}, \bibinfo {author} {\bibfnamefont {A.}~\bibnamefont {Martin}},
  \bibinfo {author} {\bibfnamefont {A.}~\bibnamefont {Reinisch}},\ and\
  \bibinfo {author} {\bibfnamefont {J.}~\bibnamefont {Gao}},\ }\bibfield
  {title} {\bibinfo {title} {New {Method} for {Determining} the {Quality}
  {Factor} and {Resonance} {Frequency} of {Superconducting}
  {Micro}-{Resonators} from {Sonnet} {Simulations}},\ }\href
  {https://doi.org/10.1007/s10909-014-1099-3} {\bibfield  {journal} {\bibinfo
  {journal} {Journal of Low Temperature Physics}\ }\textbf {\bibinfo {volume}
  {176}},\ \bibinfo {pages} {538} (\bibinfo {year} {2014})}\BibitemShut
  {NoStop}%
\bibitem [{\citenamefont {Blais}\ \emph {et~al.}(2004)\citenamefont {Blais},
  \citenamefont {Huang}, \citenamefont {Wallraff}, \citenamefont {Girvin},\
  and\ \citenamefont {Schoelkopf}}]{blais_cavity_2004}%
  \BibitemOpen
  \bibfield  {author} {\bibinfo {author} {\bibfnamefont {A.}~\bibnamefont
  {Blais}}, \bibinfo {author} {\bibfnamefont {R.-S.}\ \bibnamefont {Huang}},
  \bibinfo {author} {\bibfnamefont {A.}~\bibnamefont {Wallraff}}, \bibinfo
  {author} {\bibfnamefont {S.~M.}\ \bibnamefont {Girvin}},\ and\ \bibinfo
  {author} {\bibfnamefont {R.~J.}\ \bibnamefont {Schoelkopf}},\ }\bibfield
  {title} {\bibinfo {title} {Cavity quantum electrodynamics for superconducting
  electrical circuits: {An} architecture for quantum computation},\ }\href
  {https://doi.org/10.1103/PhysRevA.69.062320} {\bibfield  {journal} {\bibinfo
  {journal} {Physical Review A}\ }\textbf {\bibinfo {volume} {69}},\ \bibinfo
  {pages} {062320} (\bibinfo {year} {2004})}\BibitemShut {NoStop}%
\bibitem [{\citenamefont {Wallraff}\ \emph {et~al.}(2004)\citenamefont
  {Wallraff}, \citenamefont {Schuster}, \citenamefont {Blais}, \citenamefont
  {Frunzio}, \citenamefont {Huang}, \citenamefont {Majer}, \citenamefont
  {Kumar}, \citenamefont {Girvin},\ and\ \citenamefont
  {Schoelkopf}}]{wallraff_strong_2004}%
  \BibitemOpen
  \bibfield  {author} {\bibinfo {author} {\bibfnamefont {A.}~\bibnamefont
  {Wallraff}}, \bibinfo {author} {\bibfnamefont {D.~I.}\ \bibnamefont
  {Schuster}}, \bibinfo {author} {\bibfnamefont {A.}~\bibnamefont {Blais}},
  \bibinfo {author} {\bibfnamefont {L.}~\bibnamefont {Frunzio}}, \bibinfo
  {author} {\bibfnamefont {R.-S.}\ \bibnamefont {Huang}}, \bibinfo {author}
  {\bibfnamefont {J.}~\bibnamefont {Majer}}, \bibinfo {author} {\bibfnamefont
  {S.}~\bibnamefont {Kumar}}, \bibinfo {author} {\bibfnamefont {S.~M.}\
  \bibnamefont {Girvin}},\ and\ \bibinfo {author} {\bibfnamefont {R.~J.}\
  \bibnamefont {Schoelkopf}},\ }\bibfield  {title} {\bibinfo {title} {Strong
  coupling of a single photon to a superconducting qubit using circuit quantum
  electrodynamics},\ }\href {https://doi.org/10.1038/nature02851} {\bibfield
  {journal} {\bibinfo  {journal} {Nature}\ }\textbf {\bibinfo {volume} {431}},\
  \bibinfo {pages} {162} (\bibinfo {year} {2004})}\BibitemShut {NoStop}%
\bibitem [{\citenamefont {Koch}\ \emph {et~al.}(2007)\citenamefont {Koch},
  \citenamefont {Yu}, \citenamefont {Gambetta}, \citenamefont {Houck},
  \citenamefont {Schuster}, \citenamefont {Majer}, \citenamefont {Blais},
  \citenamefont {Devoret}, \citenamefont {Girvin},\ and\ \citenamefont
  {Schoelkopf}}]{koch_charge-insensitive_2007}%
  \BibitemOpen
  \bibfield  {author} {\bibinfo {author} {\bibfnamefont {J.}~\bibnamefont
  {Koch}}, \bibinfo {author} {\bibfnamefont {T.~M.}\ \bibnamefont {Yu}},
  \bibinfo {author} {\bibfnamefont {J.}~\bibnamefont {Gambetta}}, \bibinfo
  {author} {\bibfnamefont {A.~A.}\ \bibnamefont {Houck}}, \bibinfo {author}
  {\bibfnamefont {D.~I.}\ \bibnamefont {Schuster}}, \bibinfo {author}
  {\bibfnamefont {J.}~\bibnamefont {Majer}}, \bibinfo {author} {\bibfnamefont
  {A.}~\bibnamefont {Blais}}, \bibinfo {author} {\bibfnamefont {M.~H.}\
  \bibnamefont {Devoret}}, \bibinfo {author} {\bibfnamefont {S.~M.}\
  \bibnamefont {Girvin}},\ and\ \bibinfo {author} {\bibfnamefont {R.~J.}\
  \bibnamefont {Schoelkopf}},\ }\bibfield  {title} {\bibinfo {title}
  {Charge-insensitive qubit design derived from the {Cooper} pair box},\ }\href
  {https://doi.org/10.1103/PhysRevA.76.042319} {\bibfield  {journal} {\bibinfo
  {journal} {Physical Review A}\ }\textbf {\bibinfo {volume} {76}},\ \bibinfo
  {pages} {042319} (\bibinfo {year} {2007})}\BibitemShut {NoStop}%
\bibitem [{\citenamefont {Stefanazzi}\ \emph {et~al.}(2022)\citenamefont
  {Stefanazzi}, \citenamefont {Treptow}, \citenamefont {Wilcer}, \citenamefont
  {Stoughton}, \citenamefont {Bradford}, \citenamefont {Uemura}, \citenamefont
  {Zorzetti}, \citenamefont {Montella}, \citenamefont {Cancelo}, \citenamefont
  {Sussman}, \citenamefont {Houck}, \citenamefont {Saxena}, \citenamefont
  {Arnaldi}, \citenamefont {Agrawal}, \citenamefont {Zhang}, \citenamefont
  {Ding},\ and\ \citenamefont {Schuster}}]{stefanazzi_qick_2022}%
  \BibitemOpen
  \bibfield  {author} {\bibinfo {author} {\bibfnamefont {L.}~\bibnamefont
  {Stefanazzi}}, \bibinfo {author} {\bibfnamefont {K.}~\bibnamefont {Treptow}},
  \bibinfo {author} {\bibfnamefont {N.}~\bibnamefont {Wilcer}}, \bibinfo
  {author} {\bibfnamefont {C.}~\bibnamefont {Stoughton}}, \bibinfo {author}
  {\bibfnamefont {C.}~\bibnamefont {Bradford}}, \bibinfo {author}
  {\bibfnamefont {S.}~\bibnamefont {Uemura}}, \bibinfo {author} {\bibfnamefont
  {S.}~\bibnamefont {Zorzetti}}, \bibinfo {author} {\bibfnamefont
  {S.}~\bibnamefont {Montella}}, \bibinfo {author} {\bibfnamefont
  {G.}~\bibnamefont {Cancelo}}, \bibinfo {author} {\bibfnamefont
  {S.}~\bibnamefont {Sussman}}, \bibinfo {author} {\bibfnamefont
  {A.}~\bibnamefont {Houck}}, \bibinfo {author} {\bibfnamefont
  {S.}~\bibnamefont {Saxena}}, \bibinfo {author} {\bibfnamefont
  {H.}~\bibnamefont {Arnaldi}}, \bibinfo {author} {\bibfnamefont
  {A.}~\bibnamefont {Agrawal}}, \bibinfo {author} {\bibfnamefont
  {H.}~\bibnamefont {Zhang}}, \bibinfo {author} {\bibfnamefont
  {C.}~\bibnamefont {Ding}},\ and\ \bibinfo {author} {\bibfnamefont {D.~I.}\
  \bibnamefont {Schuster}},\ }\bibfield  {title} {\bibinfo {title} {The {QICK}
  ({Quantum} {Instrumentation} {Control} {Kit}): {Readout} and control for
  qubits and detectors},\ }\href {https://doi.org/10.1063/5.0076249} {\bibfield
   {journal} {\bibinfo  {journal} {Review of Scientific Instruments}\ }\textbf
  {\bibinfo {volume} {93}},\ \bibinfo {pages} {044709} (\bibinfo {year}
  {2022})}\BibitemShut {NoStop}%
\bibitem [{\citenamefont {Newcomb}(1966)}]{newcomb_linear_1966}%
  \BibitemOpen
  \bibfield  {author} {\bibinfo {author} {\bibfnamefont {R.~W.}\ \bibnamefont
  {Newcomb}},\ }\href@noop {} {\emph {\bibinfo {title} {Linear {Multiport}
  {Synthesis}}}}\ (\bibinfo  {publisher} {McGraw-Hill},\ \bibinfo {year}
  {1966})\BibitemShut {NoStop}%
\bibitem [{\citenamefont {Foster}(1924)}]{foster_reactance_1924}%
  \BibitemOpen
  \bibfield  {author} {\bibinfo {author} {\bibfnamefont {R.~M.}\ \bibnamefont
  {Foster}},\ }\bibfield  {title} {\bibinfo {title} {A reactance theorem},\
  }\href@noop {} {\bibfield  {journal} {\bibinfo  {journal} {Bell System
  technical journal}\ }\textbf {\bibinfo {volume} {3}},\ \bibinfo {pages} {259}
  (\bibinfo {year} {1924})}\BibitemShut {NoStop}%
\bibitem [{\citenamefont {Nigg}\ \emph {et~al.}(2012)\citenamefont {Nigg},
  \citenamefont {Paik}, \citenamefont {Vlastakis}, \citenamefont {Kirchmair},
  \citenamefont {Shankar}, \citenamefont {Frunzio}, \citenamefont {Devoret},
  \citenamefont {Schoelkopf},\ and\ \citenamefont
  {Girvin}}]{nigg_black-box_2012}%
  \BibitemOpen
  \bibfield  {author} {\bibinfo {author} {\bibfnamefont {S.~E.}\ \bibnamefont
  {Nigg}}, \bibinfo {author} {\bibfnamefont {H.}~\bibnamefont {Paik}}, \bibinfo
  {author} {\bibfnamefont {B.}~\bibnamefont {Vlastakis}}, \bibinfo {author}
  {\bibfnamefont {G.}~\bibnamefont {Kirchmair}}, \bibinfo {author}
  {\bibfnamefont {S.}~\bibnamefont {Shankar}}, \bibinfo {author} {\bibfnamefont
  {L.}~\bibnamefont {Frunzio}}, \bibinfo {author} {\bibfnamefont {M.~H.}\
  \bibnamefont {Devoret}}, \bibinfo {author} {\bibfnamefont {R.~J.}\
  \bibnamefont {Schoelkopf}},\ and\ \bibinfo {author} {\bibfnamefont {S.~M.}\
  \bibnamefont {Girvin}},\ }\bibfield  {title} {\bibinfo {title} {Black-{Box}
  {Superconducting} {Circuit} {Quantization}},\ }\href
  {https://doi.org/10.1103/PhysRevLett.108.240502} {\bibfield  {journal}
  {\bibinfo  {journal} {Physical Review Letters}\ }\textbf {\bibinfo {volume}
  {108}},\ \bibinfo {pages} {240502} (\bibinfo {year} {2012})}\BibitemShut
  {NoStop}%
\bibitem [{\citenamefont {Taubin}(1991)}]{taubin_estimation_1991}%
  \BibitemOpen
  \bibfield  {author} {\bibinfo {author} {\bibfnamefont {G.}~\bibnamefont
  {Taubin}},\ }\bibfield  {title} {\bibinfo {title} {Estimation of planar
  curves, surfaces, and nonplanar space curves defined by implicit equations
  with applications to edge and range image segmentation},\ }\href
  {https://doi.org/10.1109/34.103273} {\bibfield  {journal} {\bibinfo
  {journal} {IEEE Transactions on Pattern Analysis and Machine Intelligence}\
  }\textbf {\bibinfo {volume} {13}},\ \bibinfo {pages} {1115} (\bibinfo {year}
  {1991})},\ \bibinfo {note} {conference Name: IEEE Transactions on Pattern
  Analysis and Machine Intelligence}\BibitemShut {NoStop}%
\bibitem [{\citenamefont {Probst}\ \emph {et~al.}(2015)\citenamefont {Probst},
  \citenamefont {Song}, \citenamefont {Bushev}, \citenamefont {Ustinov},\ and\
  \citenamefont {Weides}}]{probst_efficient_2015}%
  \BibitemOpen
  \bibfield  {author} {\bibinfo {author} {\bibfnamefont {S.}~\bibnamefont
  {Probst}}, \bibinfo {author} {\bibfnamefont {F.~B.}\ \bibnamefont {Song}},
  \bibinfo {author} {\bibfnamefont {P.~A.}\ \bibnamefont {Bushev}}, \bibinfo
  {author} {\bibfnamefont {A.~V.}\ \bibnamefont {Ustinov}},\ and\ \bibinfo
  {author} {\bibfnamefont {M.}~\bibnamefont {Weides}},\ }\bibfield  {title}
  {\bibinfo {title} {Efficient and robust analysis of complex scattering data
  under noise in microwave resonators},\ }\href
  {https://doi.org/10.1063/1.4907935} {\bibfield  {journal} {\bibinfo
  {journal} {Review of Scientific Instruments}\ }\textbf {\bibinfo {volume}
  {86}},\ \bibinfo {pages} {024706} (\bibinfo {year} {2015})}\BibitemShut
  {NoStop}%
\end{thebibliography}%

\end{document}